\documentclass[conference]{IEEEtran}
\usepackage{amsmath}
\usepackage{amssymb}
\usepackage{amsfonts}
\usepackage{graphicx}

\newtheorem{theorem}{Theorem}

\newtheorem{algorithm}[theorem]{Algorithm}

\newtheorem{problem}[theorem]{Problem}

\newtheorem{remark}[theorem]{Remark}

\numberwithin{equation}{section}
\numberwithin{theorem}{section}
\input{tcilatex}
\begin{document}

\title{Computing the Matched Filter in Linear Time\\
\textit{\ }{\small To Solomon Golomb for the occasion of his 80 birthday
mazel tov}}

\author{\authorblockN{Alexander Fish}
\authorblockA{Department of Mathematics\\
University of Wisconsin\\
Madison, WI 53706, USA\\
Email: afish@math.wisc.edu}
\and
\authorblockN{Shamgar Gurevich}
\authorblockA{Department of Mathematics\\
University of Wisconsin\\
Madison, WI 53706, USA\\
Email: shamgar@math.wisc.edu}
\and
\authorblockN{Ronny Hadani}
\authorblockA{Department of Mathematics\\
University of Texas\\
Austin,  TX 78712, USA\\
Email: hadani@math.utexas.edu}
\and
\and
\authorblockN{Akbar Sayeed}
\authorblockA{Department of Electrical Engineering\\
University of Wisconsin\\
Madison, WI 53706, USA\\
Email: akbar@engr.wisc.edu}
\and
\authorblockN{Oded Schwartz}
\authorblockA{Department of Computer Science\\ 
University of California\\
Berkeley, CA 94720, USA\\
Email: odedsc@eecs.berkeley.edu}
}%

\maketitle%

\begin{abstract}%

A fundamental problem in wireless communication is the \textit{%
time-frequency shift} (TFS) problem: Find the time-frequency shift of a
signal in a noisy environment. The shift is the result of time
asynchronization of a sender with a receiver, and of non-zero speed of a
sender with respect to a receiver. A classical solution of a discrete analog
of the TFS problem is called the \textit{matched filter} algorithm. It uses
a pseudo-random waveform $S(t)$ of the length $p,$ and its arithemtic
complexity is $O(p^{2}\cdot \log (p))$, using fast Fourier transform. In
these notes we introduce a novel approach of designing new waveforms that
allow faster matched filter algorithm. We use techniques from group
representation theory to design waveforms $S(t),$ which enable us to
introduce two \textit{fast matched filter} (FMF) algorithms, called the 
\textit{flag algorithm}, and the \textit{cross algorithm.} These methods
solve the TFS problem in $O(p\cdot \log (p))$ operations. We discuss
applications of the algorithms to mobile communication, GPS, and radar.

\end{abstract}%

\section{\textbf{Introduction}}

Denote by $\mathcal{H}=%
\mathbb{C}
(\mathbb{F}_{p})$ the vector space of complex valued functions on the finite
field $\mathbb{F}_{p}=\{0,1,...,p-1\},$ where addition and multiplication is
done modulo the odd prime number $p.$ The vector space $\mathcal{H}$ is
equipped with the standard inner product $\left\langle
f_{1},f_{2}\right\rangle =\tsum\limits_{t\in \mathbb{F}_{p}}f_{1}(t)%
\overline{f_{2}(t)}$, for $f_{1},f_{2}\in \mathcal{H}$, and will be referred
to as the \textit{Hilbert space of digital signals}.

Let us start with a motivational problem.

\subsection{\textbf{Mobile communication problem}}

We consider the following mathematical model of mobile communication \cite%
{TV}. There exists a collection of users $j=1,...,r$, each holding a bit $%
b_{j}\in \{\pm 1\},$ and a private signal $S_{j}\in \mathcal{H}$. User $j$
transmits its message $b_{j}\cdot S_{j}$ to a base station (antenna), and
the base station receives the superposition sum 
\begin{equation}
R(t)=\sum_{j=1}^{r}b_{j}\cdot e^{\frac{2\pi i}{p}\omega _{j}\cdot t}\cdot
S_{j}(t+\tau _{j})\text{ }+\mathcal{W}(t),\text{ \ \ }t\in \mathbb{F}_{p},
\label{RMC}
\end{equation}%
where $\mathcal{W}\in \mathcal{H}$ denotes a random white noise of mean
zero, $\tau _{j}$ encodes the time asynchronization of user $j$ with the
base station, $\omega _{j}$ encodes the radial velocity of user $j$ with
respect to the base station, and $i=\sqrt{-1}.$

The base station "knows" the signals $S_{j}$'s and $R$. The objective
is:\medskip\ 

\begin{problem}[Mobile communication problem]
\label{MC}Extract the bits $b_{j},$ $j=1,...,r.$ \medskip
\end{problem}

A resolution of Problem \ref{MC} will be deduced (see Section \ref{SMC})
from our solution to the following problem.

\subsection{\textbf{The time-frequency shift (TFS) problem}}

We have $r$ signals $S_{j}\in \mathcal{H}$, $j=1,...,r,$ called the \textit{%
sender} waveforms. Additionally, we are given the \textit{receiver} waveform 
$R\in \mathcal{H}$, which satisfies 
\begin{equation}
R(t)=\sum_{j=1}^{r}e^{\frac{2\pi i}{p}\omega _{j}\cdot t}\cdot S_{j}(t+\tau
_{j})\text{ }+\mathcal{W}(t),\text{ \ \ }t\in \mathbb{F}_{p},  \label{RTFS}
\end{equation}%
where $\mathcal{W}\in \mathcal{H}$ denotes a random white noise of mean
zero, and $\left( \tau _{j},\omega _{j}\right) \in \mathbb{F}_{p}\times 
\mathbb{F}_{p}$, $j=1,...,r.$ We will call the pairs $\left( \tau
_{j},\omega _{j}\right) $ the \textit{time-frequency shifts}, and the vector
space $V=\mathbb{F}_{p}\times \mathbb{F}_{p}$ the\ \textit{time-frequency
plane}.

The precise formulation of the \textit{time-frequency shift problem} is the
following:\medskip

\begin{problem}[TFS problem]
\label{TFS}Given the waveforms $S_{j},$ $j=1,...,r,$ and $R,$ extract the
time-frequency shifts $\left( \tau _{j},\omega _{j}\right) \in V$, $%
j=1,...,r.$ \medskip
\end{problem}

\subsection{\textbf{The matched filter (MF) algorithm}}

A classical solution \cite{GG,GHS1,GHS2,HCM,TV,V,WG} to Problem \ref{TFS},
is the \textit{matched filter algorithm}. For a fixed $k\in \{1,...,r\},$ we
define the following matched filter (MF) matrix of the sender $S_{k}$, and
the receiver $R$:%
\begin{equation}
\mathcal{M}[S_{k},R](\tau ,\omega )=\left\langle e^{\frac{2\pi i}{p}\omega
\cdot t}\cdot S_{k}(t+\tau ),R(t)\right\rangle ,\ \ (\tau ,\omega )\in V.
\label{MF}
\end{equation}%
A direct verification shows that for $\zeta _{j}=e^{\frac{2\pi i}{p}(\tau
\omega _{j}-\omega \tau _{j})}$, $j=1,...,r$, we have 
\begin{eqnarray}
\mathcal{M}[S_{k},R](\tau ,\omega ) &=&\zeta _{k}\cdot \mathcal{M}%
[S_{k},S_{k}](\tau -\tau _{k},\omega -\omega _{k})  \label{MFI} \\
&&+\sum_{j\neq k}\zeta _{j}\cdot \mathcal{M[}S_{k},S_{j}](\tau -\tau
_{j},\omega -\omega _{j})  \notag \\
&&+O(\frac{NSR}{\sqrt{p}}),  \notag
\end{eqnarray}%
where $NSR=\frac{1}{SNR}$ is the inverse of the signal-to-noise ratio
between the waveform $S_{k}$ and $\mathcal{W}.$ For simplicity, we assume
that the $NSR$ is not too large, and, for the rest of the paper, we will
omit the last term in (\ref{MFI}).

In order to extract the time-frequency shift $(\tau _{k},\omega _{k}),$
using the matched filter, it is "standard" (see \cite%
{GG,GHS1,GHS2,HCM,TV,V,WG}) to use almost-orthogonal pseudo-random signals $%
S_{j}\in \mathcal{H}$ of norm one. Namely, all the summands in right-hand
side of (\ref{MFI}) are of size $O(\frac{1}{\sqrt{p}})$, with the exception
that for $(\tau ,\omega )=(\tau _{k},\omega _{k})$ we have $\mathcal{M}%
[S_{k},S_{k}](\tau -\tau _{k},\omega -\omega _{k})=1.$ Hence,%
\begin{equation}
\left\vert \mathcal{M}[S_{k},R](\tau ,\omega )\right\vert =\left\{ 
\begin{array}{c}
1+\varepsilon _{r,p}\text{, if }(\tau ,\omega )=(\tau _{k},\omega _{k}); \\ 
\varepsilon _{r,p}\text{, \ \ \ \ \ if }(\tau ,\omega )\neq (\tau
_{k},\omega _{k}),%
\end{array}%
\right.  \label{Peak}
\end{equation}%
where $\varepsilon _{r,p}=O(\frac{r}{\sqrt{p}}).$\bigskip

\begin{figure}[ht]
\includegraphics[clip,height=4cm]{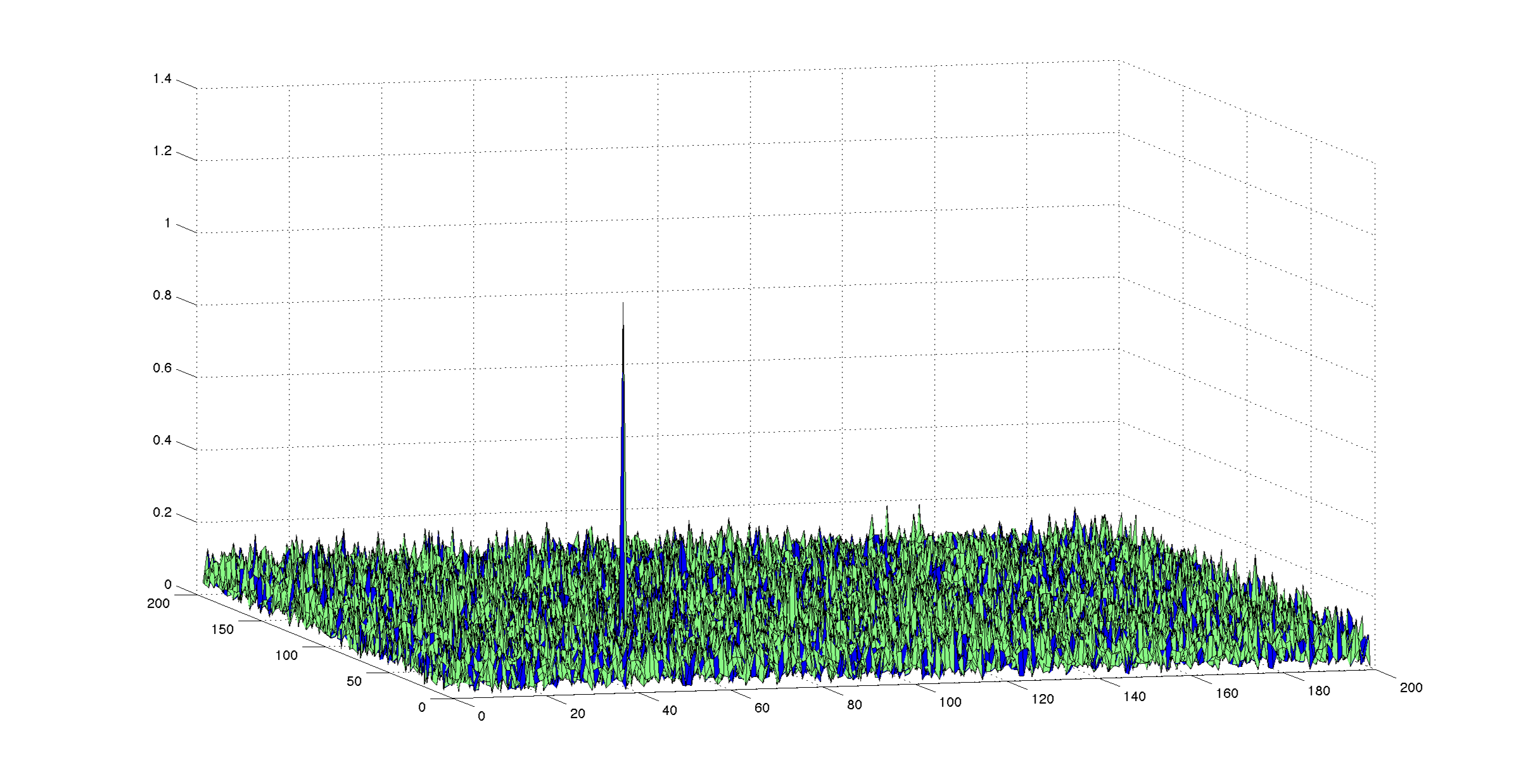}\\
\caption{$\left\vert \mathcal{M}%
[S_{1},R]\right\vert $\textit{\ with pseudo-random }$S_{1},S_{2}$\textit{,
and }$(\protect\tau _{1},\protect\omega _{1})=(50,50)$}
\end{figure}



Identity (\ref{Peak}) suggests the following "entry-by-entry" algorithmic
solution to TFS\ problem: Compute the matrix $\mathcal{M}[S_{k},R],$ and
choose $(\tau _{k},\omega _{k})$ for which $\left\vert \mathcal{M}%
[S_{k},R](\tau _{k},\omega _{k})\right\vert \approx 1.$ However, this
solution of TFS problem is very expensive in terms of arithmetic complexity,
i.e., the number of arithmetic (multiplication, and addition) is $O(r\cdot
p^{3}).$ One can do better using a "line-by-line" computation. This is due
to the next observation.\smallskip

\begin{remark}[FFT]
\label{FFT}The restriction of the matrix $\mathcal{M[}S_{k},R]$ to any line
(not necessarily through the origin) in the time-frequency plane $V,$ is a
convolution that can be computed, using the fast Fourier transform algorithm
(FFT), in $O(p\cdot \log (p))$ arithmetic operations.\smallskip
\end{remark}

As a consequence of Remark \ref{FFT}, one can solve TFS\ problem in $%
O(r\cdot p^{2}\cdot \log (p))$ arithmetic operations. To the best of our
knowledge, the "line-by-line" computation is also the fastest method which
exists in the literature \cite{OMB}. Note that computing one entry in $%
\mathcal{M[}S_{k},R]$ costs already $O(p)$ operations. This leads to the
following \textit{fast matched filter (FMF)} problem:\medskip

\begin{problem}[FMF problem]
\label{FMF}Design waveforms $S_{j}\in \mathcal{H}$, $j=1,...,r,$ to solve
TFS problem in almost linear time for shift.
\end{problem}

\subsection{\textbf{The flag method}}

We introduce the \textit{flag method} to propose a solution to FMF problem.
We will show how to associate with the $p+1$ lines, through $(0,0)$ in the
time-frequency plane, $L_{j},$ $j=1,...,p+1,$ a system of almost orthogonal
waveforms $S_{L_{j}}\in \mathcal{H},$ that we will call \textit{flags}. The
system satisfies 
\begin{equation}
\left\vert \mathcal{M}[S_{L_{k}},R](\tau ,\omega )\right\vert =\left\{ 
\begin{array}{c}
2+\varepsilon _{r,p}\text{, \ if }(\tau ,\omega )=(\tau _{k},\omega _{k});%
\text{ \ \ \ \ \ \ \ } \\ 
1+\varepsilon _{r,p}\text{, \ if }(\tau ,\omega )\in L_{k}^{\prime
}\smallsetminus (\tau _{k},\omega _{k}); \\ 
\varepsilon _{r,p}\text{,\ \ if }(\tau ,\omega )\in V\smallsetminus
L_{k}^{\prime },\text{\ \ \ \ \ \ \ \ \ \ \ \ \ \ \ }%
\end{array}%
\right.  \label{flag}
\end{equation}%
where $\varepsilon _{r,p}=O(\frac{r}{\sqrt{p}}),$ $R$ is the receiver
waveform (\ref{RTFS}), defined with respect to any $r$ flags containing $%
S_{L_{k}},$ and $L_{k}^{\prime }$ is the shifted line $L_{k}+(\tau
_{k},\omega _{k})$.

\bigskip

\begin{figure}[ht]
\includegraphics[clip,height=4cm]{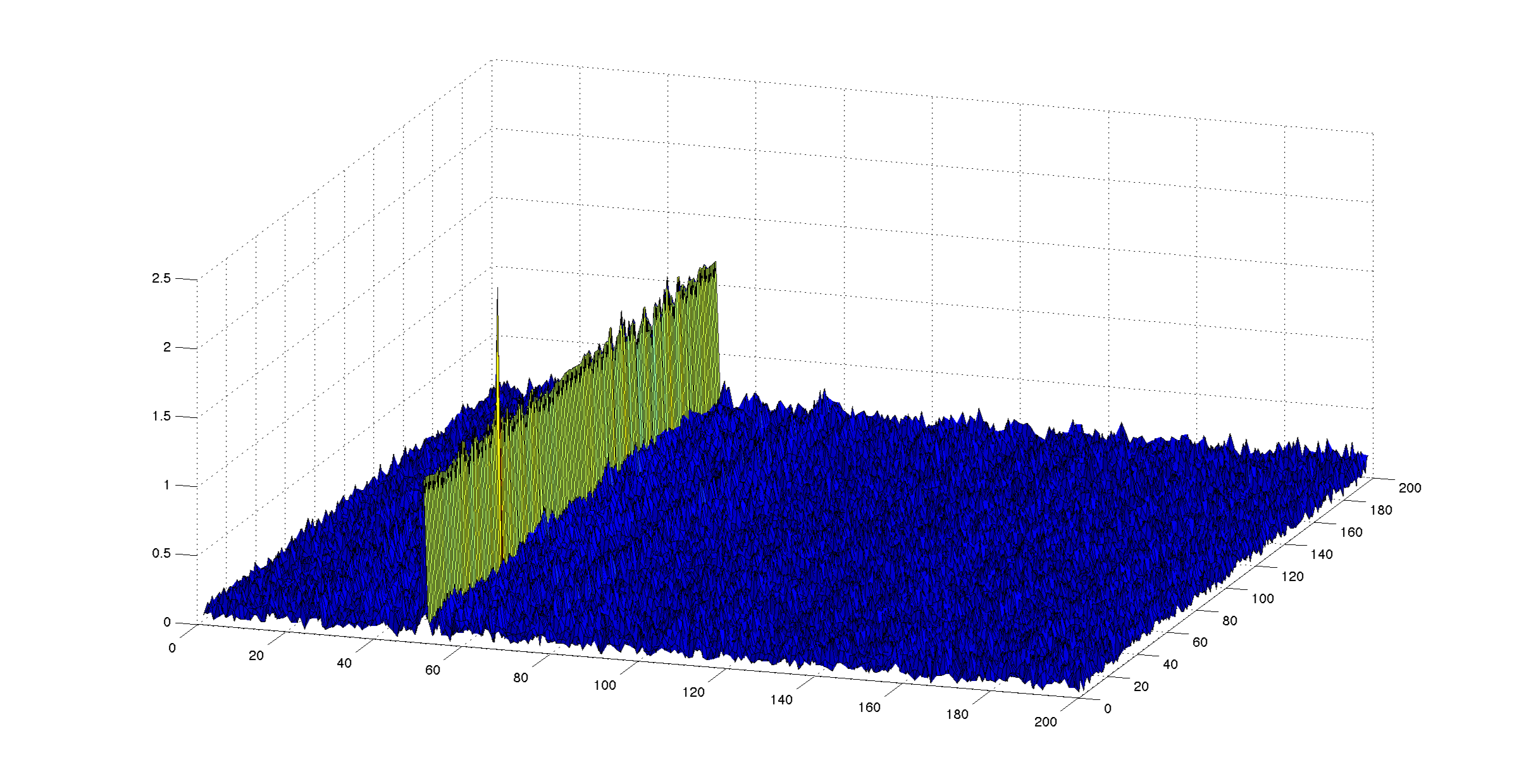}\\
\caption{$\left\vert
M[S_{L_{1}},R]\right\vert $\textit{\ with two flags }$S_{L_{1}},S_{L_{2}}$%
\textit{, and }$(\protect\tau _{1},\protect\omega _{1})=(50,50)$}
\end{figure}


Identity (\ref{flag}) suggests the "flag" algorithmic solution to FMF\
problem in the case that the number of waveforms $r\ll \sqrt{p},$ and $p$ is
sufficiently large. In the following we assume that $R$ and $S_{L_{k}}$ are
as in (\ref{flag}).\smallskip \smallskip

\begin{algorithm}[Flag algorithm]
\begin{itemize}
\item Choose a line $L_{k}^{\bot }$ different from $L_{k}.$

\item Compute $\mathcal{M}[S_{L_{k}},R]$ on $L_{k}^{\bot }$. Find $(\tau
,\omega )$ such that $\left\vert \mathcal{M}[S_{L_{k}},R](\tau ,\omega
)\right\vert $ $\approx 1$, i.e., $(\tau ,\omega )$ on the shifted line $%
L_{k}+(\tau _{k},\omega _{k}).$

\item Compute $\mathcal{M}[S_{L_{k}},R]$ on $L_{k}+(\tau _{k},\omega _{k})$
and find $(\tau ,\omega )$ such that $\left\vert \mathcal{M}%
[S_{L_{k}},R](\tau ,\omega )\right\vert \approx 2.\smallskip $
\end{itemize}
\end{algorithm}

The arithmetic complexity of the flag algorithm is $O(r\cdot p\log (p)),$
using the FFT (Remark \ref{FFT}).

\begin{figure}[ht]
\includegraphics[clip,height=4cm]{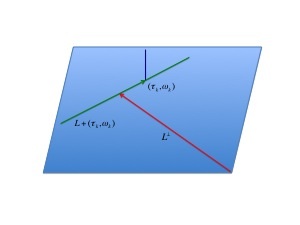}\\
\caption{Diagram of the flag algorithm}
\end{figure}


\subsection{The cross method}

Another solution to the TFS problem, and subsequently to the mobile
communication problem\textit{,} is the \textit{cross method. }The idea is
similar to the flag method, i.e., first to find a line on which the
time-frequency shift is located, and then to search on the line to find the
time-frequency shift. We will show how to associate with the $\frac{p+1}{2}$
distinct pairs of lines $L,M\subset V$ a system of almost-orthogonal
waveforms $S_{_{L},_{M}},$ that we will call \textit{crosses. }The system
satisfies%
\begin{equation*}
\left\vert \mathcal{M[}S_{_{L},_{M}},R\mathcal{]}\right\vert {\small =}%
\left\{ 
\begin{array}{c}
_{2+\varepsilon _{r,p}\text{, \ if }(\tau ,\omega )=(\tau
_{_{L},_{M}},\omega _{_{L},_{M}});}\text{ \ \ \ \ \ \ \ \ \ } \\ 
_{1+\varepsilon _{r,p}\text{, \ if }(\tau ,\omega )\in (L^{\prime }\cup
M^{\prime })\smallsetminus (\tau _{_{L},_{M}},\omega _{_{L},_{M}});} \\ 
_{\varepsilon _{r,p}\text{,\ \ if }(\tau ,\omega )\in V\smallsetminus
(L^{\prime }\cup M^{\prime }),}\text{\ \ \ \ \ \ \ \ \ \ \ \ \ \ \ \ \ }%
\end{array}%
\right.
\end{equation*}%
where $\varepsilon _{r,p}=O(\frac{r}{\sqrt{p}}),$ $R$ is the receiver
waveform (\ref{RTFS}), defined with respect to any $r$ different crosses
containing $S_{_{L},_{M}},$ and $L^{\prime }=L+(\tau _{_{L},_{M}},\omega
_{_{L},_{M}})$, $M^{\prime }=M+(\tau _{_{L},_{M}},\omega _{_{L},_{M}}).$

\begin{figure}[ht]
\includegraphics[clip,height=4cm]{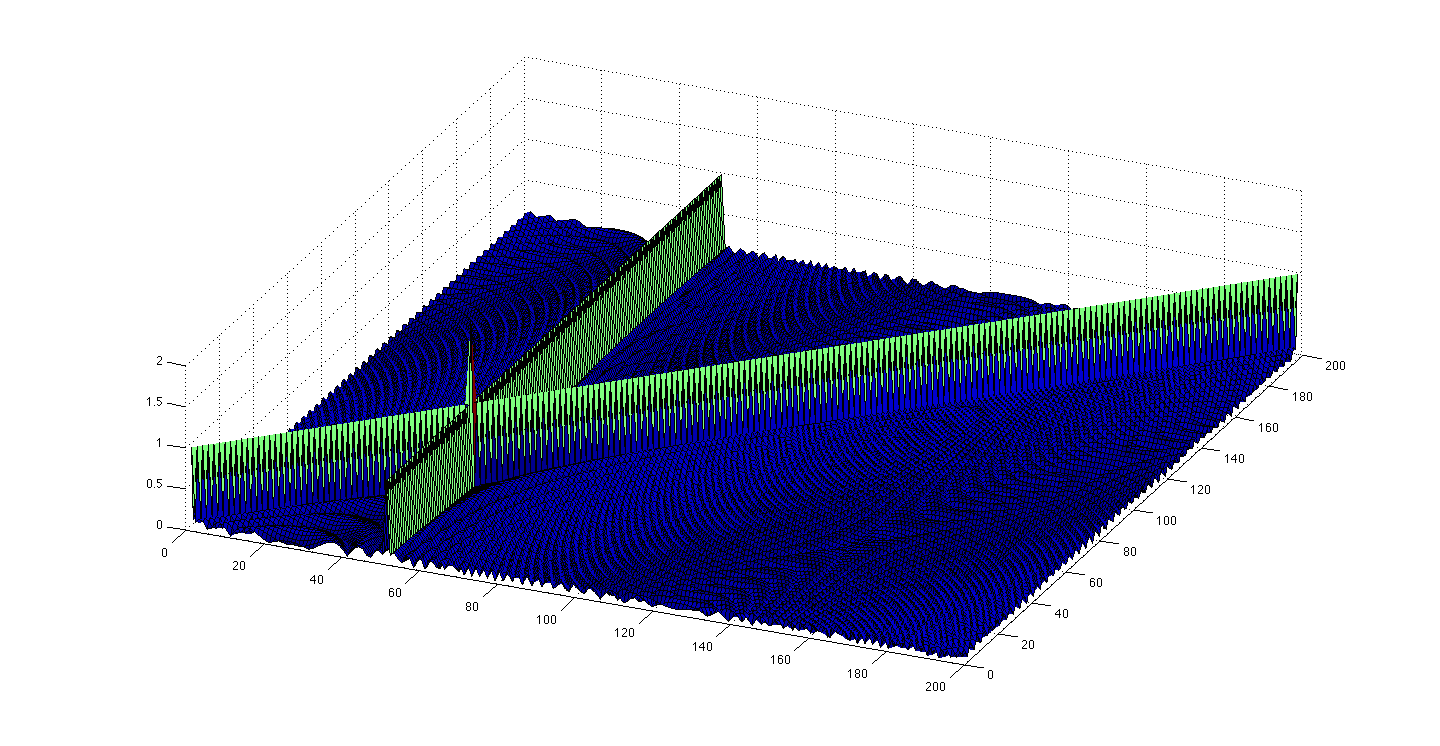}\\
\caption{$|\mathcal{M}%
[S_{_{L_{1}},_{M_{1}}},R]|$\ with crosess $%
S_{_{L_{1}},_{M_{1}}},S_{_{L_{2}},_{M_{2}}}$, and $(\protect\tau _{1},%
\protect\omega _{1})=(50,50)$}
\end{figure}


The arithmetic complexity of the cross method is $O(r\cdot p\log (p)),$
using the FFT (Remark \ref{FFT}).

\subsection{\textbf{Solution to the mobile communication problem\label{SMC}}}

Looking back to Problem \ref{MC}, we see that the flag and cross algorithms
suggest a fast $O(r\cdot p\cdot \log (p))$ solution to extract ALL the bits $%
b_{k}.$ Indeed, identity (\ref{MFI}) implies that $b_{k}\approx \mathcal{M}%
[S_{k},R](\tau _{k},\omega _{k})/2,$ \ $k=1,...,r,$ where $R$ is the
waveform (\ref{RMC}), with $S_{j}=S_{L_{j}}$, $j=1,...,r$, for the flag
method, or $S_{j}=S_{_{L_{j}},_{M_{j}}}$, $j=1,...,r$, for the cross method,
and $r\ll \sqrt{p}.$

\section{\textbf{The Heisenberg--Weil flag system}}

The flag waveforms, that play the main role in the flag algorithm, are of a
special form. \ Each of them is a sum of a pseudorandom signal and a
structural signal. The first has the MF matrix which is almost delta
function at the origin, and the MF matrix of the second is supported on a
line. The designs of these waveforms are done using group representation
theory. The pseudorandom signals are designed \cite{GHS1, GHS2, WG} using
the Weil representation, and will be called Weil (peak) signals\footnote{%
For the purpose of the Flag method, other pseudorandom signals may work.}.
The structural signals are designeded \cite{H, HCM} using the Heisenberg
representation, and will be called Heisenberg (lines) signals. We will call
the collection of all flag waveforms, the Heisenberg--Weil flag system. In
this section we briefly recall constructions, and properties of these
waveforms. A more comprehensive treatment, including proofs, will appear in 
\cite{FGHSS}.

\subsection{\textbf{The Heisenberg (lines) system\label{HS}}}

Consider the following collection of unitary operators, called Heisenberg
operators, that act on the Hilbert space of digital signals:%
\begin{equation}
\left\{ 
\begin{array}{c}
\pi (\tau ,\omega ):\mathcal{H\rightarrow \mathcal{H}},\text{ \ }\tau
,\omega \in \mathbb{F}_{p}; \\ 
\pi (\tau ,\omega )=\mathrm{M}_{\omega }\circ \mathrm{L}_{\tau },%
\end{array}%
\right. \text{ \ \ \ \ \ \ \ \ \ \ \ \ \ \ }  \label{Heis}
\end{equation}%
where $\mathrm{L}_{\tau }[f](t)=f(t+\tau )$ is the time-shift operator, $%
\mathrm{M}_{\omega }[f](t)=e^{\frac{2\pi i}{p}\omega \cdot t}\cdot f(t)$ is
the frequency-shift operator, for every $f\in \mathcal{H}$, $t\in \mathbb{F}%
_{p},$ and $\circ $ denotes composition of operators.

The operators (\ref{Heis}) do not commute in general, but rather obey the
Heisenberg commutation relations $\pi (\tau ,\omega )\circ \pi (\tau
^{\prime },\omega ^{\prime })=e^{\frac{2\pi i}{p}(\tau \omega ^{\prime
}-\omega \tau ^{\prime })}\cdot \pi (\tau ^{\prime },\omega ^{\prime })\circ
\pi (\tau ,\omega ).$ The expression $\tau \omega ^{\prime }-\omega \tau
^{\prime }$ vanishes if $(\tau ,\omega )$, $(\tau ^{\prime },\omega ^{\prime
})$ belong to the same line. Hence, for a given line $L\subset V=\mathbb{F}%
_{p}\times \mathbb{F}_{p}$ we have a commutative collection of unitary
operators%
\begin{equation}
\pi (\ell ):\mathcal{H\rightarrow \mathcal{H}},\text{ }\ell \in L.
\label{piL}
\end{equation}%
We use the theorem from linear algebra about simultaneous diagonalization of
commuting unitary operators, and obtain \cite{H, HCM} a natural orthonormal
basis $\mathcal{B}_{L\text{ }}\subset \mathcal{H}$ consisting of common
eigenfunctions for all the operators (\ref{piL}). The system of all such
bases $\mathcal{B}_{L},$ where $L$ runs over all lines through the origin in 
$V,$ will be called the \textit{Heisenberg (lines) system}. We will need the
following result \cite{H, HCM}:\smallskip

\begin{theorem}
\label{HT}The Heisenberg system satisfies the properties

\begin{enumerate}
\item \textit{Line. }For every line $L\subset V$, and every $f_{L}\in 
\mathcal{B}_{L}$, we have 
\begin{equation*}
\left\vert \mathcal{M}[f_{L},f_{L}](\tau ,\omega )\right\vert =\left\{ 
\begin{array}{c}
1,\text{ if }\left( \tau ,\omega \right) \in L; \\ 
0,\text{ if }\left( \tau ,\omega \right) \notin L.%
\end{array}%
\right.
\end{equation*}

\item \textit{Almost-orthogonality.} For every two lines $L_{1}\neq
L_{2}\subset V$, and every $f_{L_{1}}\in \mathcal{B}_{L_{1}}$, $f_{L_{2}}\in 
\mathcal{B}_{L_{2}},$ we have%
\begin{equation*}
\left\vert \mathcal{M}[f_{L_{1}},f_{L_{2}}](\tau ,\omega )\right\vert =\frac{%
1}{\sqrt{p}},\text{ \ }
\end{equation*}%
for every $(\tau ,\omega )\in V.$
\end{enumerate}
\end{theorem}

\begin{figure}[ht]
\includegraphics[clip,height=4cm]{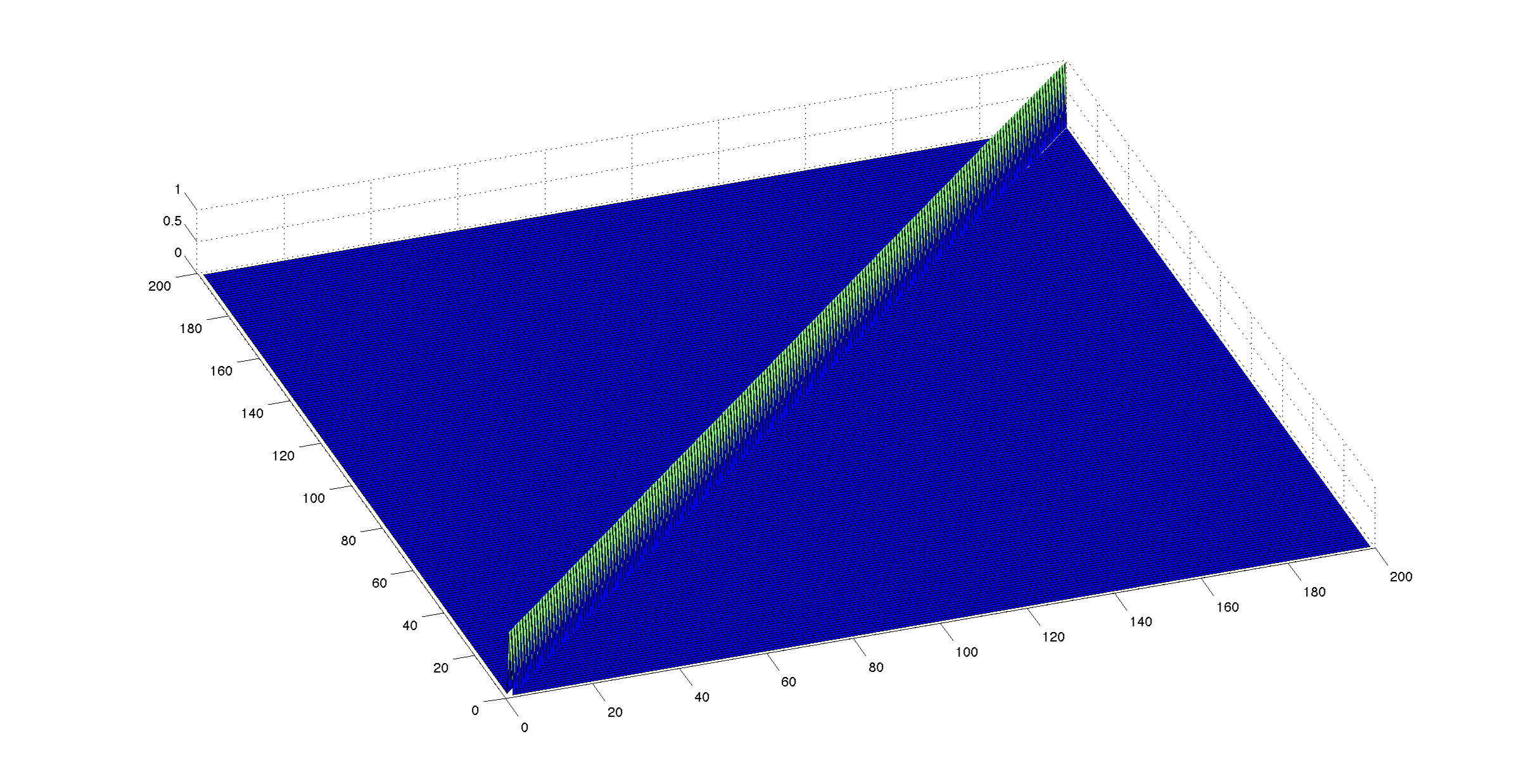}\\
\caption{$\left\vert \mathcal{M}%
[f_{L},f_{L}]\right\vert $ for $L=\{(\protect\tau ,\protect\tau );$ $\protect%
\tau \in \mathbb{F}_{p}\}$}
\end{figure}


\subsection{\textbf{The Weil (peaks) system}}

Consider the following collection of matrices 
\begin{equation*}
G=SL_{2}(\mathbb{F}_{p})=\left\{ 
\begin{pmatrix}
a & b \\ 
c & d%
\end{pmatrix}%
;\text{ }a,b,c,d\in \mathbb{F}_{p},\text{ and }ad-bc=1\right\} .
\end{equation*}%
Note that $G$ is in a natural way a \textit{group} \cite{A} with respect to
the operation of matrix multiplication. It is called the \textit{special
linear group} of order two over $\mathbb{F}_{p}.$ Each element $g\in G$ acts
on the time-frequency plane $V$ via the change of coordinates $v\mapsto
g\cdot v.$ For every $g\in G$, let $\rho (g)$ be a linear operator on $%
\mathcal{H}$ which is a solution of the following system of $p^{2}$ linear
equations:%
\begin{equation}
\Sigma _{g}:\text{ }\rho (g)\circ \pi (\tau ,\omega )=\pi (g\cdot (\tau
,\omega ))\circ \rho (g),\text{\ \ }\tau ,\omega \in \mathbb{F}_{p},
\label{S}
\end{equation}%
where $\pi $ is defined by (\ref{Heis}). Denote by $\mathrm{Sol}(\Sigma
_{g}) $ the space of all solutions to System (\ref{S}). The following is a
basic result \cite{W}:

\begin{theorem}[Stone--von Neumann--Schur-Weil]
\label{S-vN}There exist a unique collection of solutions $\{\rho (g)\in 
\mathrm{Sol}(\Sigma _{g});$ $g\in G\},$ which are unitary operators, and
satisfy the homomorphism condition $\rho (g\cdot h)=\rho (g)\circ \rho
(h).\smallskip $
\end{theorem}

Denote by $U(\mathcal{H)}$ the collection of all unitary operators on the
Hilbert space of digital signals $\mathcal{H}$. Theorem \ref{S-vN}
establishes the map $\rho :G\rightarrow U(\mathcal{H})$, which is called the 
\textit{Weil representation }\cite{W}\textit{. }The group $G$ is not
commutative, but contains a special class of maximal commutative subgroups
called tori\footnote{%
There are order of $p^{2}$ tori in $SL_{2}(\mathbb{F}_{p}).$} \cite{GHS1,
GHS2}. Each torus $T\subset G$ acts via the Weil representation operators 
\begin{equation}
\rho (g):\mathcal{H\rightarrow \mathcal{H}}\text{, \ }g\in T.  \label{W}
\end{equation}%
This is a commutative collection of diagonalizable operators, and it admits 
\cite{GHS1, GHS2} a natural orthonormal basis $\mathcal{B}_{T}$ for $%
\mathcal{H}$, consisting of common eigenfunctions. The system of all such
bases $\mathcal{B}_{T},$ where $T$ runs over all tori in $G,$ will be called
the \textit{Weil (peaks) system}. We will need the following result \cite%
{GHS1, GHS2}:\smallskip

\begin{theorem}
\label{WPeak}The Weil system satisfies the properties

\begin{enumerate}
\item \textit{Peak. }For every torus $T\subset G$, and every $\varphi
_{T}\in \mathcal{B}_{T}$, we have%
\begin{equation*}
\left\vert \mathcal{M}[\varphi _{T},\varphi _{T}](\tau ,\omega )\right\vert
=\left\{ 
\begin{array}{c}
1,\text{ if }\left( \tau ,\omega \right) =(0,0); \\ 
\leq \frac{2}{\sqrt{p}},\text{ if }\left( \tau ,\omega \right) \neq (0,0).%
\end{array}%
\right.
\end{equation*}

\item \textit{Almost-orthogonality.} For every two tori $T_{1},$ $%
T_{2}\subset G$, and every $\varphi _{T_{1}}\in \mathcal{B}_{T_{1}}$, $%
\varphi _{T_{2}}\in \mathcal{B}_{T_{2}},$ with $\varphi _{T_{1}}\neq \varphi
_{T_{2}},$ we have%
\begin{equation*}
\left\vert \mathcal{M}[\varphi _{T_{1}},\varphi _{T_{2}}](\tau ,\omega
)\right\vert \leq \left\{ 
\begin{array}{c}
\frac{4}{\sqrt{p}},\text{ if }T_{1}\neq T_{2}; \\ 
\frac{2}{\sqrt{p}},\text{ if }T_{1}=T_{2},%
\end{array}%
\right. \text{ \ }
\end{equation*}%
for every $(\tau ,\omega )\in V.$
\end{enumerate}
\end{theorem}

\begin{figure}[ht]
\includegraphics[clip,height=4cm]{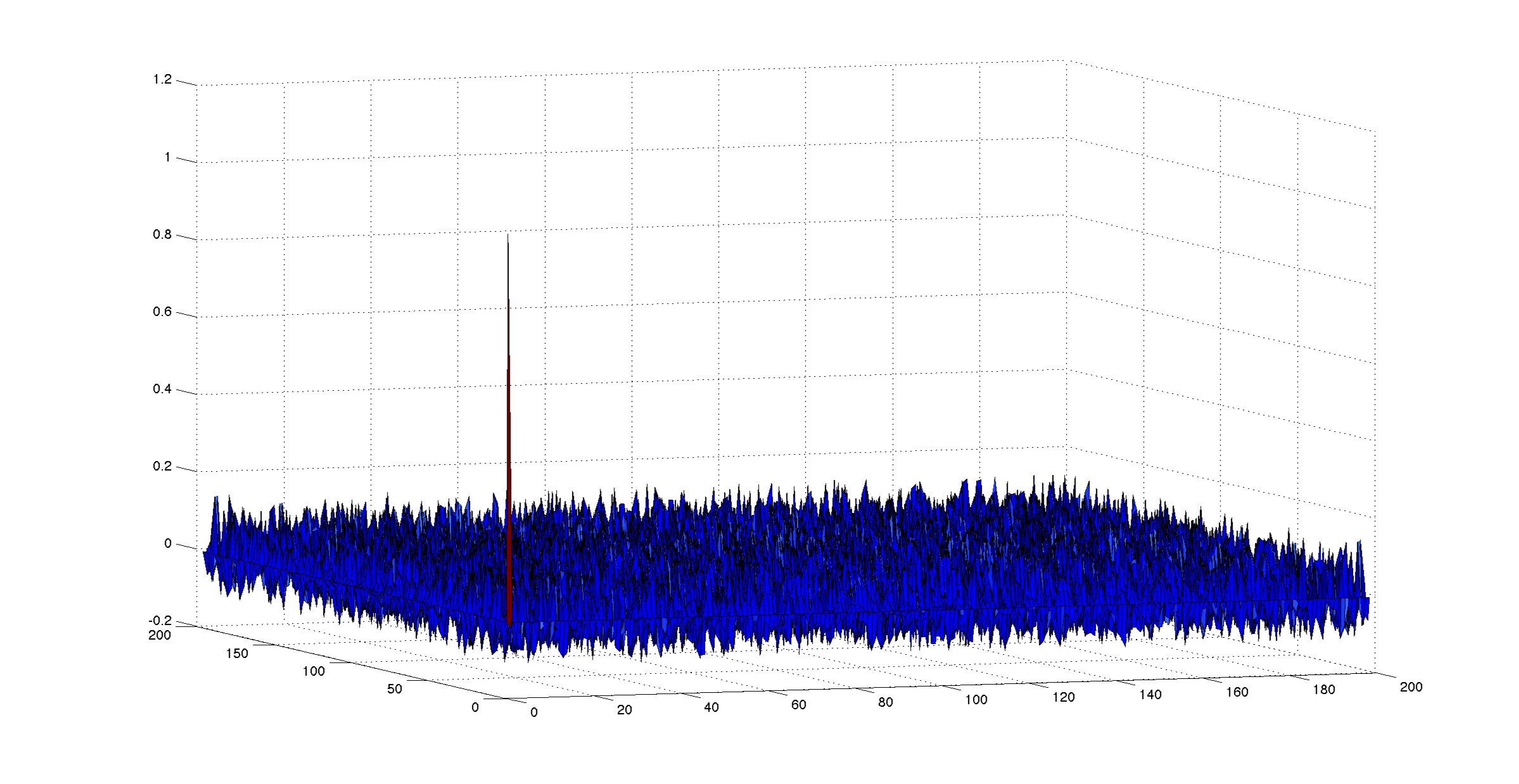}\\
\caption{$\mathcal{M}[\protect\varphi %
_{T},\protect\varphi _{T}]$ for $T=\{%
\protect\begin{pmatrix}
a & 0\protect \\ 
0 & a^{-1}%
\protect\end{pmatrix}%
;0\neq a\in \mathbb{F}_{p}\}$ }
\end{figure}


\subsection{\textbf{The Heisenberg--Weil system}}

We define the \textit{Heisenberg--Weil system} of waveforms. This is the
collection of signals in $\mathcal{H}$, which are of the form $%
S_{L}=f_{L}+\varphi _{T}$, where $f_{L}$ and $\varphi _{T}$ are Heisenberg
and Weil waveforms, respectively. The main technical result of this paper is:

\begin{theorem}
\label{HW}The Heisenberg--Weil system satisfies the properties

\begin{enumerate}
\item \textit{Flag. }For every line $L\subset V$, torus $T\subset G$, and
every flag $S_{L}=f_{L}+\varphi _{T},$ with $f_{L}\in \mathcal{B}_{L},$ $%
\varphi _{T}\in \mathcal{B}_{T}$, we have 
\begin{equation*}
\left\vert \mathcal{M}[S_{L},S_{L}](\tau ,\omega )\right\vert =\left\{ 
\begin{array}{c}
2+\epsilon _{p}\text{, \ if }(\tau ,\omega )=(0,0);\text{ \ \ \ \ \ \ \ } \\ 
1+\varepsilon _{p}\text{, \ if }(\tau ,\omega )\in L\smallsetminus (0,0);%
\text{ \ \ } \\ 
\text{ }\varepsilon _{p}\text{, \ if }(\tau ,\omega )\in V\smallsetminus L,%
\text{\ \ \ \ \ \ \ \ \ \ \ \ \ \ \ \ }%
\end{array}%
\right.
\end{equation*}%
where $|\epsilon _{p}|\leq \frac{4}{\sqrt{p}},$ and $|\varepsilon _{p}|\leq 
\frac{6}{\sqrt{p}}.$

\item \textit{Almost-orthogonality.} For every two lines $L_{1}\neq
L_{2}\subset V$, tori $T_{1},$ $T_{2}\subset G,$ and every two flags $%
S_{L_{j}}=f_{L_{j}}+\varphi _{T_{j}},$ with $f_{L_{j}}\in \mathcal{B}%
_{L_{j}} $, $\varphi _{T_{j}}\in \mathcal{B}_{T_{j}}$, $j=1,2,$ $\varphi
_{T_{1}}\neq \varphi _{T_{2}},$ we have 
\begin{equation*}
\left\vert \mathcal{M}[S_{L_{1}},S_{L_{2}}](\tau ,\omega )\right\vert \leq
\left\{ 
\begin{array}{c}
\frac{9}{\sqrt{p}},\text{ if }T_{1}\neq T_{2}; \\ 
\frac{7}{\sqrt{p}},\text{ if }T_{1}=T_{2},%
\end{array}%
\right. \text{ \ }
\end{equation*}%
for every $(\tau ,\omega )\in V.$
\end{enumerate}
\end{theorem}

A proof of Theorem \ref{HW} will appear in \cite{FGHSS}.

\begin{figure}[ht]
\includegraphics[clip,height=4cm]{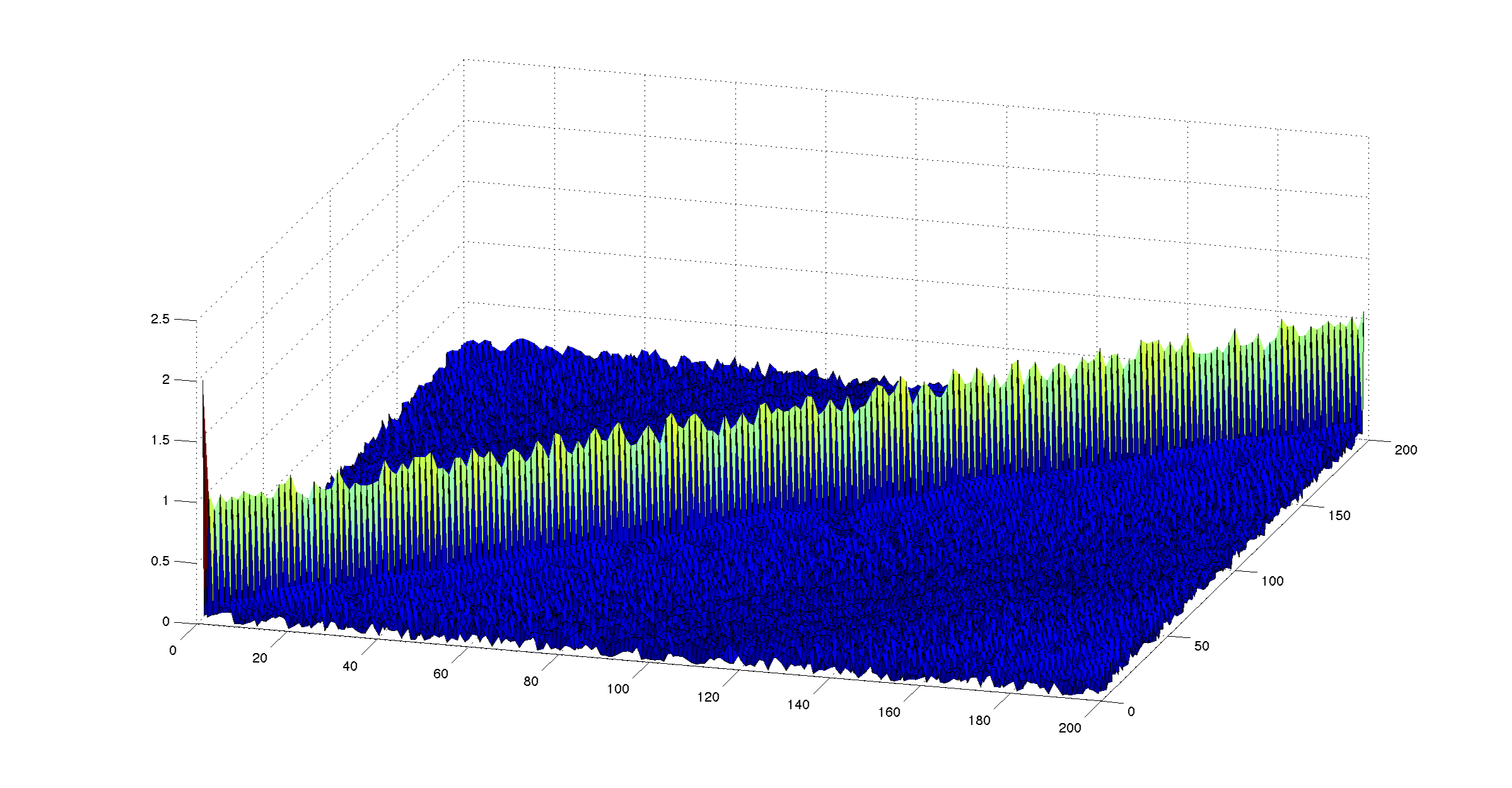}\\
\caption{$\left\vert \mathcal{M}[S_{L},S_{L}]\right\vert $ for
Heisenberg--Weil flag with $L=\{(\protect\tau ,\protect\tau );$ $\protect%
\tau \in \mathbb{F}_{p}\}$ }
\end{figure}

\smallskip

\begin{remark}
As a consequence of Theorem \ref{HW} we obtain families of $p+1$
almost-orthogonal flag waveforms which can be used for solving the TFS and
mobile communication problems in almost linear time.
\end{remark}

\section{\textbf{The Heisenberg cross system}}

We define the \textit{Heisenberg cross system} of waveforms. This is the
collection of signals in $\mathcal{H}$, which are of the form $%
S_{_{L},_{M}}=f_{L}+f_{M}$, where $f_{L},f_{M},$ $L\neq M,$ are Heisenberg
waveforms defined in Section \ref{HS}. The following follows immediately
from Theorem \ref{HT}:

\begin{theorem}
\label{HC}The Heisenberg cross system satisfies the properties

\begin{enumerate}
\item \textit{Cross. }For every pair of distinct lines $L,M\subset V$, and
every cross $S_{_{L},_{M}}=f_{L}+f_{M}$, with $f_{L}\in \mathcal{B}_{L},$ $%
f_{M}\in \mathcal{B}_{M}$, we have%
\begin{equation*}
\left\vert \mathcal{M}[S_{_{L},_{M}},S_{_{L},_{M}}](\tau ,\omega
)\right\vert =\left\{ 
\begin{array}{c}
_{2+\varepsilon _{p}\text{, \ if }(\tau ,\omega )=(0,0);}\text{ \ \ \ \ \ \
\ \ \ \ \ \ \ \ \ \ } \\ 
_{1+\varepsilon _{p}\text{, \ if }(\tau ,\omega )\in (L\cup M)\smallsetminus
(0,0);}\text{ \ \ \ \ \ \ \ } \\ 
\text{ }_{\varepsilon _{p}\text{, \ if }(\tau ,\omega )\in V\smallsetminus
(L\cup M),}\text{\ \ \ \ \ \ \ \ \ \ \ \ \ \ \ }%
\end{array}%
\right.
\end{equation*}%
where $|\varepsilon _{p}|\leq \frac{2}{\sqrt{p}}.$

\item \textit{Almost-orthogonality.} For every four distinct lines $%
L_{1},M_{1},L_{2},M_{2}\subset V$, and every two crosses $%
S_{_{L_{j}},_{M_{j}}}=f_{L_{j}}+f_{M_{j}},$ $j=1,2,$ we have 
\begin{equation*}
\left\vert \mathcal{M}[S_{_{L_{1}},_{M_{1}}},S_{_{L_{2}},_{M_{2}}}](\tau
,\omega )\right\vert \leq \frac{4}{\sqrt{p}}.\text{\ }
\end{equation*}%
for every $(\tau ,\omega )\in V.$
\end{enumerate}
\end{theorem}

\begin{remark}
As a consequence of Theorem \ref{HC} we obtain families of $\frac{p+1}{2}$
almost-orthogonal cross waveforms which can be used for solving the TFS and
mobile communication problems in almost linear time.
\end{remark}

\section{\textbf{Applications to GPS and radar}}

In the introduction we described application of flag and cross methods to
mobile communication. In this section we demonstrate applications to global
positioning system (GPS), and discrete radar.

\subsection{\textbf{Application to global positioning system (GPS)}}

The model of GPS works as follows \cite{K}. A client on the earth surface
wants to know his geographical location. Satellites $j=1,...,r$ send to
earth their location. For simplicity, the location of satellite $j$ is a bit 
$b_{j}\in \{\pm 1\}.$ Satellite $j$ transmits to the earth its signal $%
S_{j}\in \mathcal{H}$ multiplied by its location $b_{j}.$ The client
receives the signal 
\begin{equation*}
R(t)=\sum_{j=1}^{r}b_{j}\cdot e^{\frac{2\pi i}{p}\omega _{j}\cdot t}\cdot
S_{j}(t+\tau _{j})\text{ }+\mathcal{W}(t),
\end{equation*}%
where $\omega _{j}$ encodes the radial velocity of satellite $j$ with
respect to the client, $\tau _{j}$ encodes the distance between satellite $j$
and the client\footnote{%
From the $\tau _{j}$ we can find \cite{K} the distance between satellite $j$
and the client, given that $r\geq 4$ and the clocks of all satellites are
synchronized.}, and $\mathcal{W}$ is a random white noise of mean
zero.\smallskip

\begin{problem}[GPS problem]
\bigskip Find $(b_{j},\tau _{j}),$ $j=1,...,r.$ \smallskip
\end{problem}

By using Heisenberg--Weil or Heisenberg cross waveforms we find the pairs $%
(b_{j},\tau _{j})$ in $O(r\cdot p\log (p))$ arithmetical operations.

\subsection{\textbf{Application to discrete radar}}

The model of discrete radar works as follows \cite{HCM}. A radar sends a
waveform $S\in \mathcal{H}$ which bounds back by $r$ targets. The signal $%
R\in \mathcal{H}$ which is received as an echo has the form\footnote{%
In practice there are intensity coefficients $0\leq \alpha _{j}\leq 1$ such
that $R(t)=\sum_{j=1}^{r}\alpha _{j}\cdot e^{\frac{2\pi i}{p}\omega
_{j}\cdot t}\cdot S(t+\tau _{j})$ $+\mathcal{W}(t).$ Assuming that $\alpha
_{j}$'s are sufficiently large our methods are applicable verbatim.}%
\begin{equation*}
R(t)=\sum_{j=1}^{r}e^{\frac{2\pi i}{p}\omega _{j}\cdot t}\cdot S(t+\tau _{j})%
\text{ }+\mathcal{W}(t),
\end{equation*}%
where $\omega _{j}$ encodes the radial velocity of target $j$ with respect
to the radar, $\tau _{j}$ encodes the distance between target $j$ and the
radar, and $\mathcal{W}$ is a random white noise of mean zero.\smallskip
\smallskip

\begin{problem}[Discrete radar problem]
Find $(\tau _{j},\omega _{j}),$ $j=1,...,r.\smallskip \smallskip $
\end{problem}

By sending Heisenberg--Weil waveform $S_{L}=f_{L}+\varphi _{T}$ we get%
\footnote{%
For simplicity we assume that all the shifted lines $L+(\tau _{j},\omega
_{j})$'s are distinct. The general case is treated similarly.} 
\begin{equation*}
\left\vert \mathcal{M}[S_{L},R](\tau ,\omega )\right\vert =\left\{ 
\begin{array}{c}
2+\varepsilon _{r,p}\text{, if }(\tau ,\omega )\in \{(\tau _{j},\omega
_{j})\};\text{ \ \ \ \ \ \ \ \ \ \ \ \ \ \ \ \ \ } \\ 
1+\varepsilon _{r,p}\text{, if }(\tau ,\omega )\in L+(\tau _{j},\omega
_{j})\smallsetminus (\tau _{j},\omega _{j});\text{ \ } \\ 
\varepsilon _{r,p}\text{, otherwise, \ \ \ \ \ \ \ \ \ \ \ \ \ \ \ \ \ \ \ \
\ \ \ \ \ \ \ \ \ \ }%
\end{array}%
\right.
\end{equation*}%
where $\varepsilon _{r,p}=O(\frac{r}{\sqrt{p}}).$

This means that by using the flag algorithm we solve the radar problem in $%
O(r\cdot p\log (p))$ arithmetical operations.

\begin{figure}[ht]
\includegraphics[clip,height=4cm]{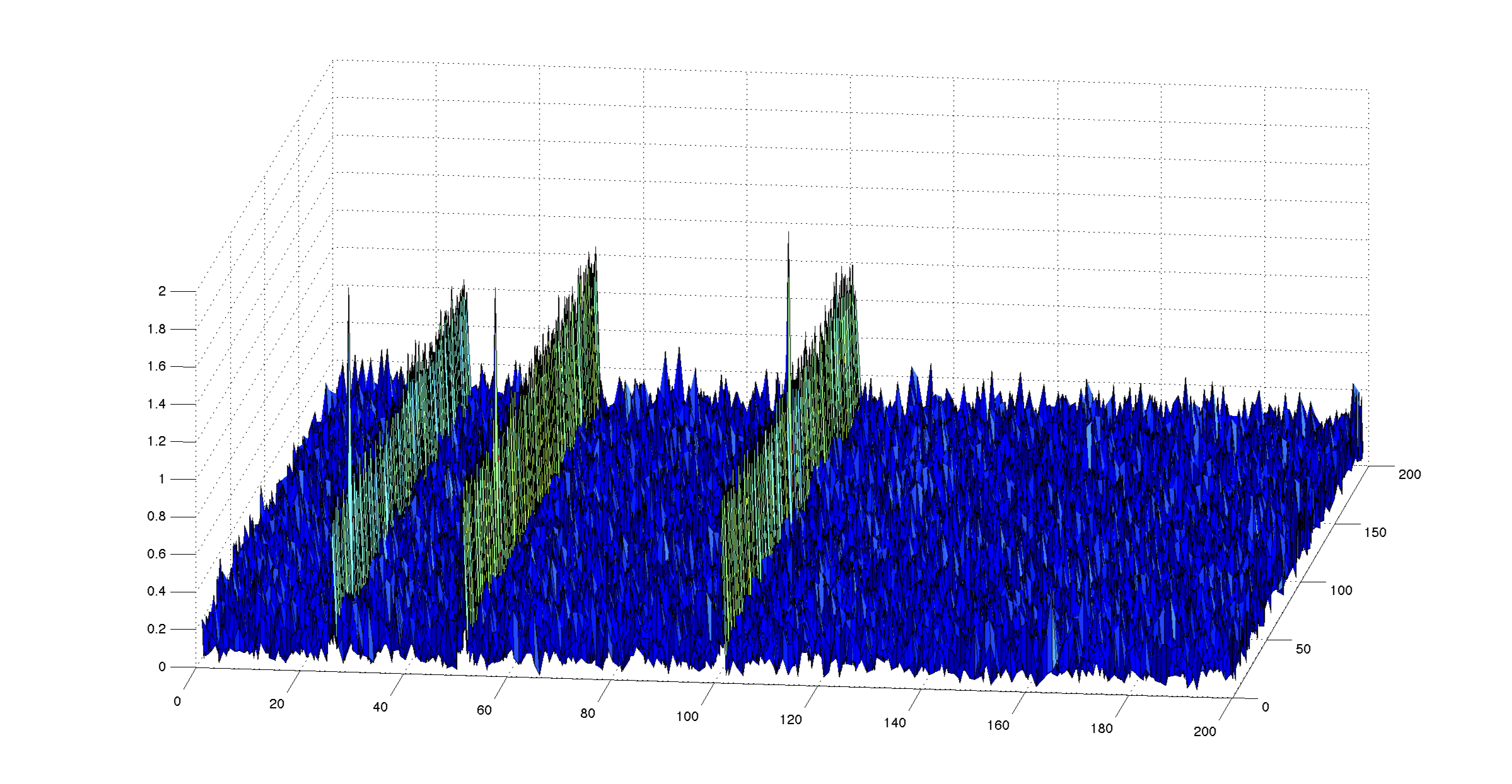}\\
\caption{$\left\vert \mathcal{M}[S_{L},R]\right\vert $ with $L=\{(%
\protect\tau ,0);$ $\protect\tau \in \mathbb{F}_{p}\}$, and shifts $\left(
50,50\right) ,$ $\left( 100,100\right) ,$ $\left( 150,150\right) $}
\end{figure}


\begin{remark}[Important]
Note that the cross method is not applicable for the discrete radar problem
if the number of targets $r>1.\smallskip $
\end{remark}

\textbf{Acknowledgement. }Warm thanks to Joseph Bernstein for his support
and encouragement in interdisciplinary research. We are grateful to Anant
Sahai, for sharing with us his thoughts, and ideas on many aspects of signal
processing and wireless communication. The project described in this paper
was initiated by a question of Mark Goresky and Andy Klapper during the
conference SETA2008, we thank them very much. We appreciate the support and
encouragement of Nigel Boston, Robert Calderbank, Solomon Golomb, Guang
Gong, Olga Holtz, Roger Howe, Peter Sarnak, Nir Sochen, and Alan Weinstein.


\begin{thebibliography}{99}
\bibitem{A} Artin M., Algebra. \textit{Prentice Hall, Inc., Englewood
Cliffs, NJ} (1991).

\bibitem{FGHSS} Fish A., Gurevich S., Hadani R., Sayeed A., Schwartz O.,
Fast matched filter and group representation theory. \textit{In \
preparation (2011).}

\bibitem{GG} Golomb, S.W. and Gong G., Signal design for good correlation.
For wireless communication, cryptography, and radar. \textit{Cambridge
University Press, Cambridge (2005).}

\bibitem{GHS1} Gurevich S., Hadani R., Sochen N., The finite harmonic
oscillator and its associated sequences. \textit{PNAS, July 22, 2008 vol.
105 no. 29 9869--9873.}

\bibitem{GHS2} Gurevich S., Hadani R., Sochen N., The finite harmonic
oscillator and its applications to sequences, communication and radar . 
\textit{IEEE Transactions on Information Theory, vol. 54, no. 9, September
2008.}

\bibitem{H} Howe R., Nice error bases, mutually unbiased bases, induced
representations, the Heisenberg group and finite geometries. \textit{Indag.
Math. (N.S.) 16 (2005), no. 3--4, 553--583.}

\bibitem{HCM} Howard, S. D., Calderbank, R., and Moran W., The finite
Heisenberg--Weyl groups in radar and communications. \textit{EURASIP J.
Appl. Signal Process (2006).}

\bibitem{K} Kaplan E., Understanding GPS Principles and Applications.\textit{%
\ Artech house, INC (1996).}

\bibitem{OMB} O'Toole J.M., Mesbah M., and Boashash B., Accurate and
efficient implementation of the time--frequency matched filter. \textit{IET
Signal Process., 2010, Vol. 4, Iss. 4, pp. 428--437.}

\bibitem{TV} Tse D., and Viswanath P., Fundamentals of Wireless
Communication. \textit{Cambridge University Press (2005).}

\bibitem{V} Verdu S., Multiuser Detection, \textit{Cambridge University
Press (1998).}

\bibitem{WG} Wang Z., and Gong G., New Sequences Design From Weil
Representation With Low Two-Dimensional Correlation in Both Time and Phase
Shifts. \textit{IEEE Transactions on Information Theory, vol. 57, no. 7,
July 2011. }

\bibitem{W} Weil A., Sur certains groupes d'operateurs unitaires. \textit{%
Acta Math. 111, 143-211 (1964).}
\end{thebibliography}
\end{document}